\documentclass[footinbib,twocolumn,showpacs,amsmath,amstex,amssymb,mathfonts,superscriptaddress,prb,longbibliography]{revtex4-1}
\pdfoutput=1
\usepackage{natbib}
\usepackage[english]{babel}
\usepackage{letltxmacro}
\usepackage{latexsym}
\LetLtxMacro{\ORIGselectlanguage}{\selectlanguage}
\makeatletter
\DeclareRobustCommand{\selectlanguage}[1]{%
  \@ifundefined{alias@\string#1}
    {\ORIGselectlanguage{#1}}
    {\begingroup\edef\x{\endgroup
       \noexpand\ORIGselectlanguage{\@nameuse{alias@#1}}}\x}%
}
\newcommand{\definelanguagealias}[2]{%
  \@namedef{alias@#1}{#2}%
}
\makeatother
\definelanguagealias{en}{english}
\definelanguagealias{English}{english}
\usepackage{graphicx}
\usepackage{amsmath}
\usepackage{amsfonts}
\usepackage{amssymb}
\usepackage{bm}
\usepackage{color}
\usepackage[percent]{overpic}
\usepackage{soul} 
\usepackage{amssymb}
\usepackage{wasysym}
\usepackage{dsfont}
\usepackage{lipsum}
\usepackage{subfigure}

\usepackage{hyperref}
\hypersetup{
    bookmarks=false,         
    unicode=false,          
    pdftoolbar=false,        
    pdfmenubar=true,        
    pdffitwindow=false,     
    pdfstartview={FitH},    
    pdftitle={},    
    pdfauthor={},     
    pdfsubject={},   
    pdfcreator={},   
    pdfproducer={}, 
    pdfkeywords={many-body localization}, 
    pdfnewwindow=true,      
    colorlinks=true,       
    linkcolor=black,          
    citecolor=blue,        
    filecolor=magenta,      
    urlcolor=blue           
}
\newcommand{\be}{\begin{equation}}
\newcommand{\ee}{\end{equation}}
\newcommand{\bea}{\begin{eqnarray}}
\newcommand{\eea}{\end{eqnarray}}

\renewcommand{\vec}[1]{{\bf #1}}

\newcommand{\ket}[1]{\left |#1 \right\rangle}
\newcommand{\bra}[1]{\left \langle #1 \right |}

\newcommand{\tsigma}{{\tilde{\sigma}}}

\newcommand*{\id}{{\normalfont\hbox{1\kern-0.15em \vrule width .8pt depth-.5pt}}}

\begin{document}

\title{Slow dynamics in translation-invariant quantum lattice models}

\author{Alexios A. Michailidis}
\affiliation{School of Physics and Astronomy, University of Leeds, Leeds, LS2 9JT, United Kingdom}
\affiliation{IST Austria, Am Campus 1, 3400 Klosterneuburg, Austria}

\author{Marko \v{Z}nidari\v{c}}
\affiliation{Physics Department, Faculty of Mathematics and Physics, University of Ljubljana, 1000 Ljubljana, Slovenia}

\author{Mariya  Medvedyeva}
\altaffiliation[Current address: ]{ASML, De Run  6501, 5504 DR, Veldhoven, The Netherlands}
\affiliation{Physics Department, Faculty of Mathematics and Physics, University of Ljubljana, 1000 Ljubljana, Slovenia}

\author{Dmitry A. Abanin}
\affiliation{Department of Theoretical Physics, University of Geneva, 24 quai Ernest-Ansermet, 1211 Geneva, Switzerland}

\author{Toma\v{z} Prosen}
\affiliation{Physics Department, Faculty of Mathematics and Physics, University of Ljubljana, 1000 Ljubljana, Slovenia}

\author{Z. Papi\'c}
\affiliation{School of Physics and Astronomy, University of Leeds, Leeds, LS2 9JT, United Kingdom}

\date{\today}
\begin{abstract}
{
Many-body quantum systems typically display fast dynamics and ballistic spreading of information. Here we address the open problem of \emph{how slow} the dynamics can be after a generic breaking of integrability by local interactions. 
We develop a method based on degenerate perturbation theory that reveals slow dynamical regimes and delocalization processes in general translation invariant models, along with accurate estimates of their delocalization time scales. Our results shed light on the fundamental questions of robustness of quantum integrable systems and the possibility of many-body localization without disorder. 
As an example, we construct a large class of one-dimensional lattice models where, despite the absence of asymptotic localization, the transient dynamics is exceptionally slow, i.e., the dynamics is indistinguishable from that of many-body localized systems for the system sizes and time scales  accessible in experiment and numerical simulations.
 }
\end{abstract}
\maketitle

\section{Introduction}

One of the central questions of quantum statistical physics is how isolated many-body systems reach thermal equilibrium. 
The process of thermalization results from the spreading of quantum information into non-local degrees of freedom during the system's unitary evolution. In ergodic systems, this spreading is fast (ballistic), since the individual eigenstates of the system are highly-entangled thermal states~\cite{DeutschETH,SrednickiETH,RigolNature}. On the other hand, there is growing interest in non-ergodic systems, which include integrable models~\cite{baxter1989} and many-body localized (MBL) systems~\cite{Anderson58,Basko06,Mirlin05}. In the latter, strong disorder significantly constrains quantum dynamics due to the emergence of an extensive number of Local Integrals of Motion (LIOMs)~\cite{Serbyn13-1, Huse13}, which cause the information to spread very slowly, i.e.,  logarithmically in time~\cite{Znidaric08,Moore12}. 

In this paper we investigate the possibility of slow dynamics in quantum lattice models with local interactions and in the absence of disorder. This is motivated by the open question of integrability breaking in quantum systems: what are the constraints on quantum dynamics following a weak but generic breaking of integrability? Does the integrability breakdown in large, translation-invariant lattice systems proceed in a smooth, classical-like KAM style~\cite{Kolmogorov}, where non-ergodic regions remain for finite integrability breaking, or is ergodicity immediately restored at asymptotic time scales~\cite{Prosen98,Prosen99}? We answer this question by providing a theoretical formalism that explains the appearance of slow ergodic dynamics on very long time scales.

For simplicity, we limit ourselves to the case of spin systems described by the Hamiltonian  $H=H_0+\lambda V$, where $H_0$ is a classical potential energy and $V$ is a quantum hopping (tunneling) term. The eigenstates of $H_0$ are classical product states of spins, while $V$ introduces quantum dynamics. Such models have recently been  investigated as potential analogs of MBL in translation-invariant systems when $\lambda$ is small~\cite{carleo2012localization,Huveneers13, Muller, QDL, SchiulazMuller15,Yao14,PhysRevB.90.165137,Garrahan15,Hickey2016,Kim2016,He2016,bols2016asymptotic,mondaini2017many,Yarloo2017,Schechter2017,1706.02603,Schiulaz}. 
Despite much  effort, the understanding of models for small but finite $\lambda$ remains less complete than that of strongly disordered systems, partly due to more pronounced finite size effects~\cite{Papic}.
For example,  while some signatures of MBL-like dynamics have been observed in such models~\cite{carleo2012localization,Muller, SchiulazMuller15,Garrahan15}, the relevance of non-perturbative effects for delocalization has also been pointed out~\cite{PhysRevB.90.165137}. Furthermore, Ref.~\onlinecite{Yao14} argued that ergodicity is restored in the thermodynamic limit, resulting in a ``quasi-MBL" phase. 

In this paper we develop a general formalism based on degenerate perturbation theory (DPT) that accurately describes the long-time dynamics of systems
where $H_0$ consists of $k\geq 2$-body interactions between particles and $V$ is a single-particle hopping term. ($H_0$ is taken to be diagonal in the computational basis and can be viewed as a model of an integrable system.) In contrast to the related perturbative arguments~\cite{PhysRevB.90.165137,SchiulazMuller15}, previously used for qualitative analysis of particular models, the DPT below is shown to yield a \emph{quantitatively} accurate description of the time evolution of the initial inhomogeneity, thus serving as a general diagnostic of the possible delocalization processes. 

The physical picture resulting from our study is that of a slow dynamical regime at intermediate time, which is exponentially long in the range of interaction terms in $H_0$. Thus, even though the studied models are asymptotically not localized, the time scales that reveal delocalization can be very large. To illustrate this slow transient dynamics, we introduce an example of a clean 1D lattice model with $k=3$-body interactions whose dynamics is shown to be indistinguishable from MBL systems at the experimentally accessible time scales. 
In contrast to previous work~\cite{Muller,Yao14,QDL,SchiulazMuller15}, the model studied here does not require very small hopping energy scales to exhibit MBL-like features, and moreover it includes only one particle species, which makes it more amenable to numerical simulations. 
This is demonstrated by matrix product state simulations on large systems ($L\lesssim 200$ particles), which provide strong numerical evidence that the 3-body model displays a clear log-like growth of entanglement entropy, in stark contrast with the integrable XXZ model. However, using DPT, we also access much longer time scales and show that the 3-body model does not display true MBL as it delocalizes in the 2nd order in DPT.  

The remainder of this paper is organized as follows. In Sec.~\ref{sec:model} we introduce the model and numerically demonstrate that it features slow dynamics.  In Sec.~\ref{sec:dpt} we present the general formalism of DPT.  In Sec.~\ref{sec:decay} we apply DPT to study the relaxation of the 3-body model, while Sec.~\ref{sec:other} contains a generalization of our results to other types of local models. Our conclusions are presented in Sec.~\ref{sec:conc}. Appendices contain further discussion on the finite-size effects of numerical simulations and details of DPT for the 3-body model.

\section{ A model with slow dynamics} \label{sec:model}

Consider the following 1D open chain of spins $\frac{1}{2}$ with length $L$:
\begin{eqnarray}
\label{Ham}
\nonumber H_0 &=& J_{1}\sum_{i=1}^{L-1}\sigma^{(3)}_{i}\sigma^{(3)}_{i+1} +J_{2}\sum_{i=1}^{L-2}\sigma^{(3)}_{i}
\sigma^{(3)}_{i+2} \\
\nonumber &+& J_{3}\sum_{i=1}^{L-2}\sigma^{(3)}_{i}\sigma^{(3)}_{i+1}
\sigma^{(3)}_{i+2}, \\
 V &=& \frac{\lambda}{2}\sum_{i=1}^{L-1}(\sigma^{+}_{i}\sigma^{-}_{i+1}+\sigma^{-}_{i}\sigma^{+}_{i+1}), 
\end{eqnarray}
where $\{\sigma_j^{(\alpha)}\}$ is the Pauli basis on site $j$ and $\sigma_j^{\pm} = (\sigma_j^{(1)}\pm i\sigma_j^{(2)})/2$. Interaction amplitudes $J_k$ are taken to be of the same order, $J$, and irrational to avoid commensurate terms in the DPT below. For numerical demonstrations we choose $J_{1} = \sqrt{2}/4$, $J_{2} = \sqrt{3}/4$, $J_{3} = \sqrt{5}/8$, but our results are not sensitive to these precise values. Moreover, as we show below, our results are insensitive to the precise value of $\lambda$, as long as $\lambda\ll J_i$.
Models like Eq.~(\ref{Ham}) physically arise in the large-$U$ limit of the Bose-Hubbard model~\cite{PachosPlenio} and in polar molecules~\cite{Buchler2007}.

We are interested in the dynamics for weak breaking of integrability, $\lambda \ll J$. We characterize dynamics by the entanglement entropy,
\begin{eqnarray}\label{eq:entropy}
S = -{\rm tr}_{A} \left( \rho_{A}\log{\rho_{A}} \right),
\end{eqnarray}
where the reduced density matrix $\rho_{A} = {\rm tr}_{B}\ket{\psi}\bra{\psi}$ is obtained by tracing out the degrees of freedom of the subsystem $B$ for a bipartition $A \cup B$ of the entire system. We perform a global quench from the product state  
\begin{eqnarray}\label{eq:psi0}
\ket{\psi}=\bigotimes_j \left( \cos{\frac{\theta_j}{2}}\ket{\downarrow_j}+e^{i \phi_j}\sin{\frac{\theta_j}{2}}\ket{\uparrow_j}\right).
\end{eqnarray}
Here $\varphi_j$ is a uniform random phase, while $\theta_j$ is obtained from a random uniform variable $\xi_j \in [-1,1]$ via the transform
\begin{eqnarray}\label{eq:cos}
\cos^r{\theta_j}=\xi_j.
\end{eqnarray}
 Parameter $r$ biases the orientation of each spin on the Bloch sphere. For $r=1$ one has spin-$1/2$ states that are random uniform on the Bloch sphere. For large $r$, on the other hand, the distribution is biased towards the poles of the Bloch sphere, with the width scaling as $\approx 1/\sqrt{r}$. In the limit $r \to \infty$ one recovers random computational states, i.e., states where each spin is either $\ket{\uparrow}$ or $\ket{\downarrow}$. Our results are independent of the choice of $r$ (see Appendix~\ref{sec:app1}) and we fix $r=11$ for optimal balance between state-to-state fluctuations and the magnitude of $S$, allowing for longest simulation times.
 
\begin{figure}[t!]
\centering
\includegraphics[width=0.9\linewidth]{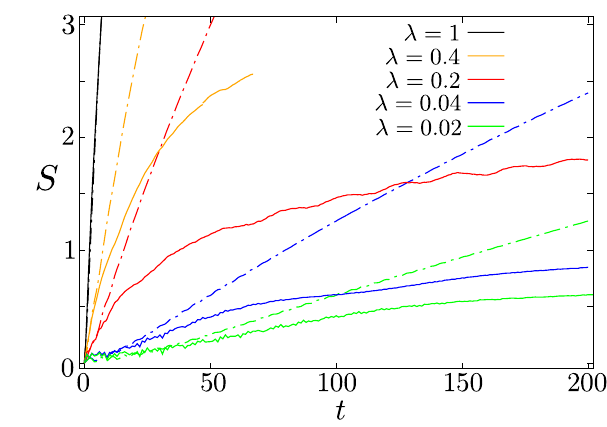}
\caption{(Color online.) Comparison of entanglement entropy growth for the XXZ model (chain curves) and the 3-body model in Eq.~(\ref{Ham}) (solid curves), for different values of $\lambda$. System size is $L=64$. The entropy growth in the 3-body model is significantly slower than for the XXZ, even for $\lambda$ as large as 0.4. At $\lambda=1$, the entropy is approximately the same in both models.}
\label{fig:lambda}
\end{figure}

In Fig.~\ref{fig:lambda} we show the representative dynamics of $S$ in the model (\ref{Ham}) starting from a single product state. Time evolution is carried out using time-evolving block-decimation (TEBD) algorithm~\citep{PhysRevLett.91.147902}. We consider a large chain of $L=64$ sites and evolve the system using bond dimensions up to $\chi=350$. We see that for all values of $\lambda$, except $\lambda=1$, there is a clear difference between the  XXZ model ($J_2=J_3=0$) and the 3-body model in Eq.~(\ref{Ham}). In particular, even for $\lambda$ as large as 0.4, we find a growth of entropy in the 3-body case which is much slower than linear. 

\begin{figure}[t]
    \includegraphics[width=\linewidth]{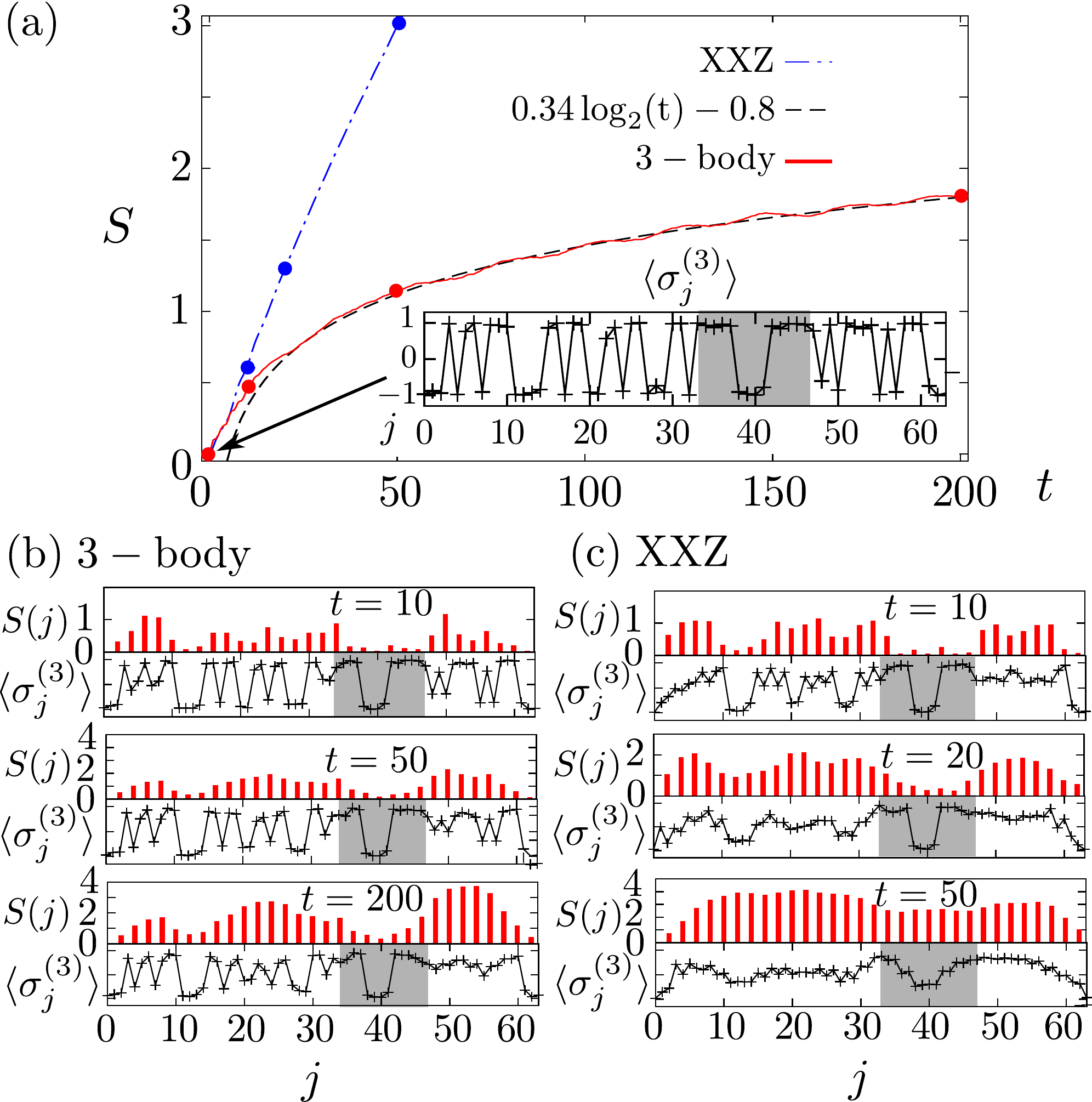}
      \vspace*{-7mm}
 \caption{ \label{Fig:big} (Color online.) (a) Slow dynamics of entanglement entropy (averaged over all cuts) for the model in Eq. (\ref{Ham}). Entropy growth can be fit with a logarithmic function of time, in contrast to the linear growth in the XXZ model (blue). Inset shows the magnetization profile of the initial state. (b) Snapshots of entropy $S(j)$ at the bond $j,j+1$ and magnetization $\langle \sigma_j^{(3)}\rangle$ at times denoted by red dots in (a). A blocking region (shaded) does not decay on the given time scale and suppresses the growth of entropy. (c) Same as (b) but for the XXZ model, where the blocking region decays by $t\sim 50$. In all simulations, $L=64$, and $\lambda=0.2$.
} 
\end{figure}

In Fig.~\ref{Fig:big}(a) we show the entropy growth for $\lambda=0.2$ for the 3-body and XXZ model, in both cases starting from the same initial state in Eq.~(\ref{eq:psi0}). On the time scales $t\lesssim 200$, we observe a clear difference between the 3-body model in Eq.~(\ref{Ham}) (red) and the XXZ model (blue). In the XXZ case, $S\propto t$, while in the 3-body case the data is consistent with $S(t)\propto \log t$. Phenomenologically, the linear growth of entropy in the XXZ model is due to the propagation of coherent quasiparticles  with a velocity  $\sim \lambda$~\cite{1742-5468-2016-6-064003,1742-5468-2005-04-P04010}. Reducing the hopping only affects the slope $S(t) \propto c(\lambda)t$, where $c(\lambda) \propto \lambda$, and the XXZ chain remains delocalized for arbitrarily small $\lambda$.  Tiny deviation from linear growth in the XXZ model is likely due to finite $L=64$ (see Appendix~\ref{sec:app1}). In contrast, the spreading of entropy in the 3-body model for small $\lambda$ appears logarithmic, even in a very large system, which is reminiscent of MBL physics~\cite{Znidaric08,Moore12}. We emphasize, however, that our numerical result in Fig.~\ref{Fig:big}  does not rule out the possibility of a power growth with a small exponent, even though the logarithmic dependence appears to give a better fit, as discussed in Appendix~\ref{sec:app1}.

To understand the mechanism of the slow dynamics, in Figs.~\ref{Fig:big}(b),(c) we examine the snapshots of entropy, evaluated at all bonds $j,j+1$, and the local magnetization $\langle \sigma_j^{(3)} \rangle$ at different times. For the given initial configuration, we observe a ``blocking region" (shaded), which does not decay on the accessible time scales in the 3-body model, but decays by the time $t\sim 50$ in the XXZ model. 

In order to understand the long-time behaviour of the model, one must go beyond the TEBD simulations (which are limited to short times) and exact diagonalization (which is susceptible to finite size effects). We therefore introduce an analytic method based on perturbation theory for the dynamics. This method will show that the model in Eq.~(\ref{Ham}) delocalizes at long (but finite) times corresponding to the 2nd order in $\lambda$. More generally, this method will allow us to understand the nature of the delocalizing processes order by order, and to quantify the role of finite-size effects. 

\section{Degenerate perturbation theory for the dynamics}\label{sec:dpt}

When $\lambda \ll J$, the Hamiltonian of Eq.~(\ref{Ham}) separates into the unperturbed part $H_{0}(\{\sigma^{(3)}\})$ and the perturbation $V$.  
Without $V$, the system is integrable -- each computational state is an eigenstate. In a semi-classical picture of small $\lambda$, the interactions still tend to localize domain walls, resulting in slow dynamics. We want to know how slow this dynamics is, and, specifically, to treat also finite values of $\lambda$. We use the Schrieffer-Wolff transformation~\cite{SchriefferWolff} (well-known in the context of ground state physics) as a framework for degenerate perturbation theory (DPT) to systematically keep track of corrections in orders of $\lambda/J$. Note that the first few orders of DPT may not capture the eigenstates at arbitrary energy. However, we show that it is possible to access dynamics in the time scales in orders of $1/\lambda$ at all energies, thereby accurately revealing the breakdown of integrability.

In this Section, we outline the DPT formalism for a general Hamiltonian of the form
\begin{equation}\label{ham1}
H = H_{0} + V,\\
\end{equation}
where we assume that $H_{0}$ is a $K$-local operator and $V$ is $R$-local, with $K>R$. We aim to find all different energy variations on the unperturbed eigenstates after applying the perturbation. 
 $H_{0}$ may contain $N$-many terms with different amplitudes $\{ J_i \}$, while $V$ contains $M$ hopping terms with amplitudes $\{ \lambda_i \}$, and we assume $J_{1},\ldots, J_{N} \gg \lambda_{1},\ldots, \lambda_{M}$. In addition, $M$ must be small enough for the perturbation to remain a correction to the unperturbed Hamiltonian.

\begin{figure}[t!]
  \centering
	\includegraphics[width=0.65\linewidth]{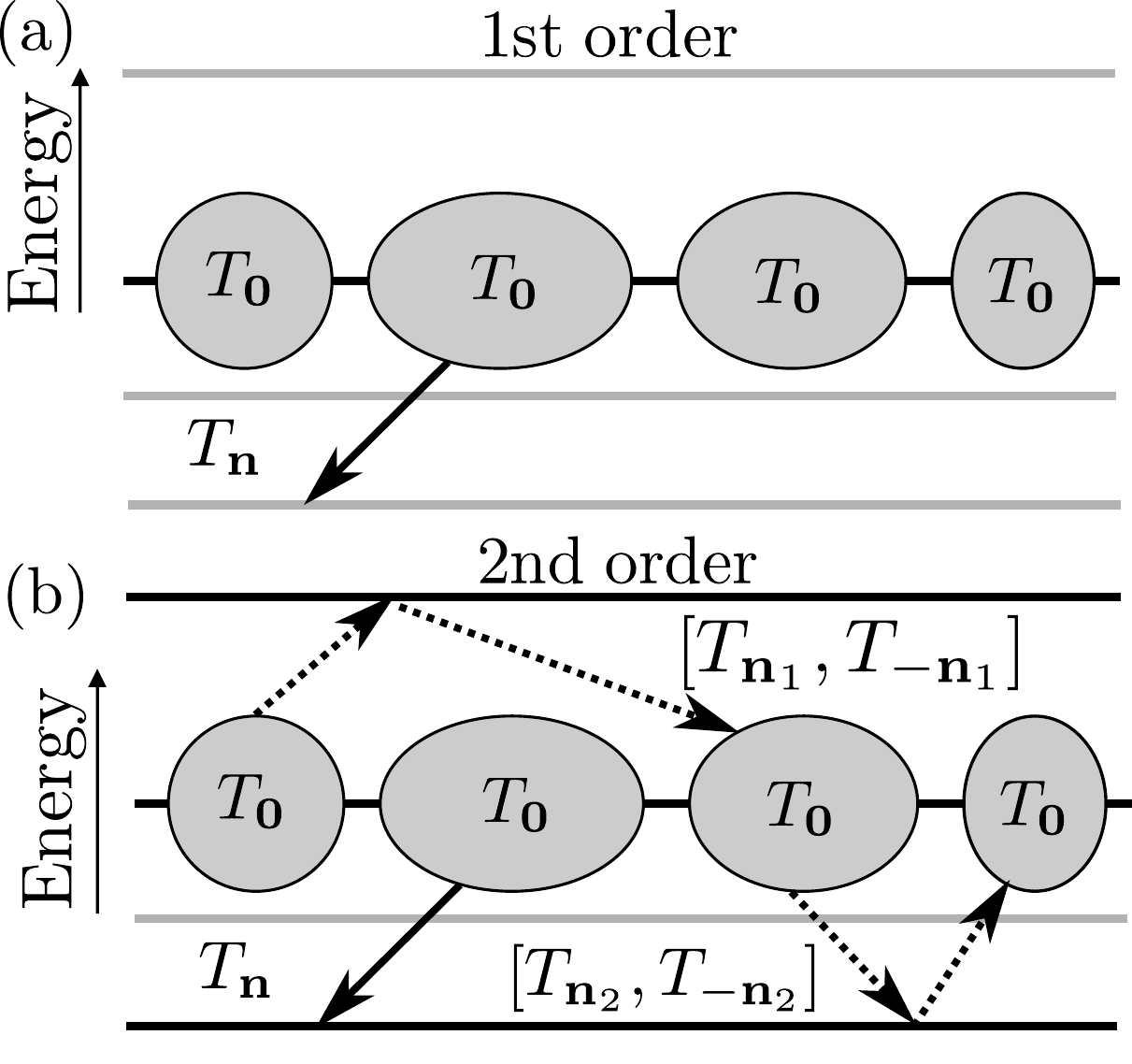}    			
 \caption{Delocalization processes in DPT. 
Solid discs represent PCDSs of varying dimensions. Solid arrows denote operations induced by the rotation of the basis. In 1st order (a), the PCDSs are disconnected. In the 2nd order (b), unblocking and resonant tunnelling (dashed arrows) connects PCDSs in larger degenerate subspaces which delocalize the system.
}
\label{fig:TRGOP1}
\end{figure}

We work in the computational $\{\sigma^{(3)}\}$ eigenbasis in which $H_0$ is diagonal. The idea of DPT is to find such a unitary transformation that will generate a block-diagonal form while eliminating higher orders in $\lambda$. To get an intuitive feeling, let us look at the effect of  $V$ on an unperturbed eigenstate $\ket{\psi}$, $H_0 \ket{\psi}=E_0 \ket{\psi}$. In the ordinary 1st order DPT, one would diagonalize $V$ in each of the subspaces ${\cal S}_0$ corresponding to a given $E_0$. As it turns out, $V$ can still have a block-diagonal structure on ${\cal S}_0$, i.e., ${\cal S}_0 = \oplus_k {\cal S}_0^{(k)}$, where subspaces ${\cal S}_0^{(k)}$ consist of degenerate states $\ket{\psi}$ connected by a single application of $V$. We call them a path-connected degenerate subspace (PCDS), see Fig~\ref{fig:TRGOP1}. In 1st order their energy $E_0$ will be split by $O(\lambda)$, but most importantly, in a given $S_0^{(k)}$ some spins have a fixed orientation for all states. This means that the dynamics happening in 1st order (on a timescale $\sim 1/\lambda$) will affect only certain spins, while others remain frozen. Those spins form ``blocking'' regions, like in Fig.~\ref{Fig:big}, and are responsible for the slow dynamics. We now formalize this reasoning and systematically extend it to higher orders.

 We start by defining operators $T$ such that $T_{\vec{n}} \subseteq V$, $\sum_{\vec{n}}T_{\vec{n}} = V$ and
\begin{equation}\label{ap:com}
\begin{aligned}
&[H_{0}, T_{\vec{n}}] = J_{\vec{n}} \; T_{\vec{n}},\\
\end{aligned}
\end{equation}
where $J_{\vec{n}}=\left(\sum_{a=1}^{N}n_{a}J_{a}\right)$. To ensure the unitarity of the transformations at any order, $T$'s have to satisfy $T^{\dagger}_{\vec{n}} = T_{-\vec{n}}$.
The operator $T_{\vec{0}}$ (if it exists) commutes with $H_{0}$ and thus spans a degenerate subspace at 1st order. Simply put, $T_{\vec{0}}$ is a projection of $V$ to the degenerate subspaces of $H_0$ (the block-diagonal part of $V$), while $T_{\vec{n}\neq \vec{0}}$ denote the corresponding off-diagonal blocks of $V$. The number of different operators $T_{\vec{n}}$ is system dependent. Each operator $T$ is translation invariant and as such it can be decomposed as
\begin{equation}
T_{\vec{n}} = \sum_{b=1}^M \lambda_{b}\sum_{i=1}^L F^{ib}_{\vec{n}}.
\end{equation}
Operators $F$ are at most ($2K+R-2$)-local and are the starting points of the expansion that we describe below. Each $F^{ib}_{\vec{n}}$ transforms a basis state into another basis state, i.e., it just flips certain spins, and the vector index $\vec{n} \equiv \{ n_a \}$ labels the difference in the unperturbed energy of a basis state $\ket{\psi}$ and the flipped one, $F^{ib}_{\vec{n}}\ket{\psi}$. Index $b$ labels different perturbations. In this work, $b$ can be omitted since we have a single perturbation and $\lambda_{1} \equiv \lambda/2$. Once $\{ F \}$ are known, any order in perturbation theory can in principle be computed. The main issue is that the support of the operators increases with respect to the order of the expansion since nested commutators are involved, thus the calculation beyond the first few orders is limited by computational resources. In Appendix~\ref{sec:dpt3b} we explicitly evaluate $F^{i}_{\vec{n}}$ for the 3-body model in Eq.~(\ref{Ham}),  and show they are 6-local operators.

To find the 1st order expansion of the Hamiltonian, a unitary transformation $\mathcal{U}^{[1]}=e^{S_{1}}$ rotates the system to the subspace where the perturbing terms that change the unperturbed energy ($\vec{n} \neq 0$) are removed, i.e., every process is resonant:
\begin{equation}\label{1storder}
\begin{aligned}
&H^{[1]} = e^{S_{1}}H e^{-S_{1}} = H+ [S_{1},H]+\ldots \\
&=H_{0} + T_{\vec{0}} + O(\lambda^2/J).
\end{aligned}
\end{equation}
For the last expression we used Eq.~(\ref{ap:com}) to pick the correct transformation 
\begin{equation}\label{1strotation}
S_{1} = \sum_{\vec{n} \neq 
\vec{0}}\frac{T_{\vec{n}}}{J_{\vec{n}}}. 
\end{equation}
For example, as shown in Appendix~\ref{sec:dpt3b} [see Eq.~\ref{ap:generators}], for the 3-body and XXZ models:
\begin{eqnarray}
\nonumber T^{\text{3-body}}_{\vec{0}} &=& \frac{\lambda}{4}\sum_{i}(\sigma^+_i \sigma^-_{i+1}+\sigma^-_i \sigma^+_{i+1})(\id+\sigma^{(3)}_{i-1}\sigma^{(3)}_{i+2})
\\
&& \times (\id+\sigma^{(3)}_{i-2}\sigma^{(3)}_{i+3})\\
T^{\text{XXZ}}_{\vec{0}} &=& \frac{\lambda}{4}\sum_{i}(\sigma^+_i \sigma^-_{i+1}+\sigma^-_i \sigma^+_{i+1})(\id+\sigma^{(3)}_{i-1}\sigma^{(3)}_{i+2}).\;\;\;\;\;
\end{eqnarray}
In 2nd order DPT, the expansion is calculated in the same spirit. The rotation removes all perturbative terms of order $O(\lambda^2/J)$. This can be calculated iteratively from the 1st order, 
\begin{equation}\label{2ndorder}
\begin{aligned}
&H^{[2]} = e^{S_{2}}e^{S_{1}}H e^{-S_{1}}e^{-S_{2}} = H+ [S_{1}+S_{2},H]+\ldots \\
&=H_{0} + T_{\vec{0}} + \sum_{\vec{n} \neq \vec{0}}\frac{1}{2J_{\vec{n}}}[T_{-\vec{n}},T_{\vec{n}}]+O(\lambda^3/J^2),
\end{aligned}
\end{equation}
where
\begin{equation}\label{2ndrotation}
S_{2} =  \sum_{\vec{n} \neq \vec{0}}\frac{1}{J^2_{\vec{n}}}[T_{\vec{n}},T_{\vec{0}}]- \sum_{\substack{\vec{n} \neq \vec{0} \\ 
\vec{n}'\neq \{-\vec{n},  \vec{0} \}
}
}
\frac{1}{2J_{\vec{n}}J_{\vec{n}+\vec{n}'}}[T_{\vec{n}},T_{\vec{n}'}].
\end{equation}

The generator of the unitary transformation of the $i$th order expansion using the iterative method is $S_{i} \sim O(\lambda^i/J^i)$. For example, the 3rd order hamiltonian is 
\begin{equation}\label{ap:3rd}
\begin{aligned}
&H^{[3]} = H^{[2]}+\sum_{
\substack{ \{ \vec{n}, \vec{n}'' \} \neq \vec{0} \\ \vec{n}'}
}
\frac{1}{2J_{\vec{n}}J_{\vec{n}''}}
[T_{\vec{n}},[T_{\vec{n}''},T_{\vec{n}'}]]+ \\ 
&- \sum_{\{\vec{n}, \vec{n}',\vec{n}''\} \neq \vec{0}}\frac{1}{6J_{\vec{n}}J_{\vec{n}'}}[T_{\vec{n}},[T_{\vec{n}'},T_{\vec{n}''}]]+O(\lambda^4/J^3).
\end{aligned}
\end{equation}
Note that the subscripts in Eq.~(\ref{ap:3rd}) also obey $\vec{n}+\vec{n}'+\vec{n}'' = \vec{0}$ in order to keep the unperturbed energy constant.

We see that the 1st order allows dynamics only within each PCDS [Fig.~\ref{fig:TRGOP1}(a)], while the 2nd order connects different PCDS through a single virtual hop [Fig.~\ref{fig:TRGOP1}(b)]. The $m$th order allows connections through $m-1$ virtual hops. The rotation of the basis consists of operators that jump between different energies and in most cases generate dephasing without transport. We note that   $[H^{[m]} ,H_{0}] = 0 $ for any order $m$ automatically follows from Eq.~(\ref{ap:com}). When we apply the DPT below, we numerically diagonalize the Hamiltonian at each order, which is a simple way to account for the splitting of the degenerate levels. 

We note that certain models, e.g., the one introduced in Ref.~\onlinecite{Garrahan15}, do not have a degenerate subspace in 1st order, i.e., $T_{\vec{0}} = \{\}$. This means that the first non-trivial order in such models is the 2nd order. In the case of Ref.~\onlinecite{Garrahan15} we only have $T_{\vec{n}_{1}},T_{-\vec{n}_{1}}$, $\vec{n}_{1} = 1$. Consequently,  not only the 1st, but also \textit{all} odd orders do not generate any new terms because odd orders include nested commutators of an odd number of $T$'s, which require that a sum of an odd number of $\vec{n}$'s must equal zero. Since the two choices are $(\vec{n}_{1},-\vec{n}_{1})$ this is impossible. Such models, which usually result from imposing classical kinetic constraints on the Hamiltonian, are expected to show the absence of relaxation for longer times due to the vanishing of odd orders of perturbation theory.

\section{Polarization decay}\label{sec:decay}

We now focus on a general dynamical probe of relaxation~\cite{PalHuse,Yao14}: we prepare an initial inhomogeneity in the spin magnetization and monitor its decay as a function of time,
\begin{equation}\label{correlator}
D(t,k)=\frac{1}{Z} \text{tr}\Big( 
e^{iHt}\tsigma^{(3)}_{k}e^{-iHt}
\tsigma^{(3)}_{k}\Big), 
\end{equation}
where we have introduced the Fourier transform of the Pauli operator
\begin{eqnarray}
\tsigma^{(3)}_{k} = \frac{1}{\sqrt{L}} \sum_{j}\sigma^{(3)}_{j}\exp(i2\pi jk/L),
\end{eqnarray}
assuming periodic boundary conditions. The normalization of $D(t,k)$ is
\begin{eqnarray}
Z=\text{tr} ( \tsigma^{(3) \dagger}_{k}\tsigma^{(3)}_{k} ),
\end{eqnarray}
where the trace is taken over zero magnetization sector. 

The interpretation of $D(t,k)$ is: throw a particle of momentum $k$ into the system; after some time remove the particle and measure the state overlap with the initial state. If the particle  scatters, the memory of the initial state gets lost and $D(t,k)\ll 1$; if the particle does not scatter, by removing it one returns to the original state and $D(t,k) = 1$. For momentum $k \approx 1$, scattering will only take place if eigenstates are extensive, thus we interpret $D(t,1)$ as a probe of delocalization of the system. Due to translation invariance, in a finite system the polarization always vanishes at $t\to\infty$. In the thermodynamic limit, if the system is in a quasi-MBL phase, one expects a time scale for the decay of  $D(t,1)$ that exponentially diverges with the system size~\cite{Yao14}.

We now apply DPT to Eq.~(\ref{correlator}). The denominator of Eq.~(\ref{correlator}) is invariant under unitary basis rotations. The time evolution operator in the numerator is transformed to the $m$th order as 
\begin{equation}
e^{-iHt} = \mathcal{U}^{[m]\dagger}e^{-iH^{[m]}t}\mathcal{U}^{[m]},
\end{equation}
where $\mathcal{U}^{[m]} = \prod_{i=1}^{m}e^{S_{m}}$. Using the cyclic property of the trace, the numerator of Eq.~(\ref{correlator}) is written as
\begin{equation}\label{ap:nom}
\text{tr}\left(\tau^{(3) *}_{k}(t)\tau^{(3)}_{k}\right),
\end{equation}
where $\tau^{(3)}_{k}=\mathcal{U}^{[m]}\tsigma^{(3)}_{k}\mathcal{U}^{[m]\dagger}$, $\tau^{(3)*}_{k}=\mathcal{U}^{[m]}\tsigma^{(3)\dagger}_{k}\mathcal{U}^{[m]\dagger}$ are the effective quasiparticles of the rotated picture. The corrections to $\tsigma^{(3)}$ due to the rotation are given in orders of $O(\lambda/J)$. The expansion of the unitary in Eq.~(\ref{ap:nom}) results in
\begin{eqnarray}
 \text{tr} \Bigl\{ e^{iH^{[m]}t}\tsigma^{(3) \dagger}_{k}e^{-iH^{[m]}t}\tsigma^{(3)}_{k} \Bigr\} + O(\lambda^2/J^2).
\end{eqnarray}
The terms of order $O(\lambda/J)$ vanish since the operators inside the trace are purely off-diagonal. To understand that, assume that the basis used to evaluate the trace is the unperturbed eigenbasis. The Hamiltonians in DPT are block-diagonal in the computational basis at any order. Every block ${\cal S}$ has some basis $\{\ket{\psi}\}$ spanned by vectors of equal unperturbed energy, 
$$
\forall \ket{\psi_{1}},\ket{\psi_{2}} \in \mathcal{S} \qquad H_{0}\ket{\psi_{1}}-H_{0}\ket{\psi_{2}} = 0.
$$ 
By default $e^{-iH^{[m]}t}$, $e^{+iH^{[m]}t}$ have the same block-diagonal structure and thus map states from ${\cal S} \rightarrow {\cal S}$. Operators  $\tsigma^{(3)\dagger}_{k}$, $\tsigma^{(3)}_{k}$ have trivial action as they are diagonal in the unperturbed eigenbasis, so they trivially map states from ${\cal S} \rightarrow {\cal S}$. On the other hand, $[S_{1},\tsigma^{(3)}_{k}]$, $[S_{1},\tsigma^{(3)\dagger}_{k}] $ always map to states outside the block, which follows trivially from the definition of $S_{1}$ in Eq.~(\ref{1strotation}). Thus an operator product which contains block-conserving and  only one of $[S_{1},\tsigma^{(3)}_{k}]$, $[S_{1},\tsigma^{(3)\dagger}_{k}]$ can only have vanishing diagonal elements. This means that magnetization decay does not feature first order basis corrections.

\subsection{Plateaus in polarization decay}

\begin{figure}[t!]
  \centering
	\includegraphics[width=0.55\linewidth]{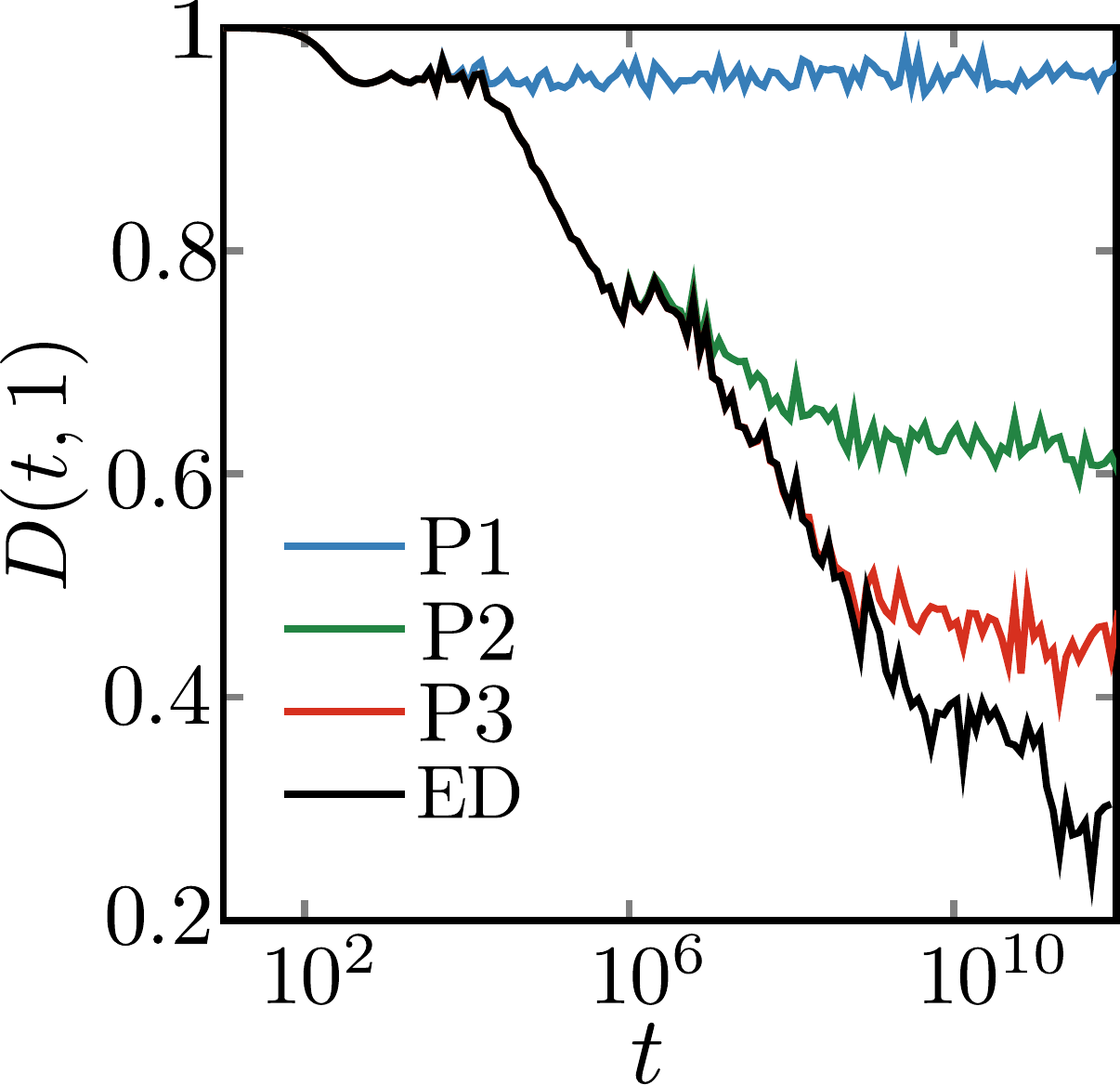}    			
 \caption{(Color online.) A comparison of the first three orders of DPT (P1-P3) against exact diagonalization. The system is described by the Hamiltonian in Eq.~(\ref{Ham}) and contains $L=14$ spins with $\lambda= 0.01$. Different orders of DPT are associated with plateaus in the decay of polarization.
}
\label{fig:TRGOP2}
\end{figure}
Using Eqs.~(\ref{1storder})-(\ref{2ndorder}), we numerically compute the magnetization  $D(t,k)$ in DPT and contrast it against exact time evolution in Fig.~\ref{fig:TRGOP2}. As explained above, in this particular calculation we can ignore basis rotations up to the time-scale to which the given order is accurate and thus model the dynamics according to  Eq.~(\ref{ap:nom}). (By contrast, the calculation of entanglement entropy would be sensitive to the basis rotations.) 

The comparison between the first three orders of DPT and exact evolution is shown in Fig.\ref{fig:TRGOP2}. Evidently, DPT is practically exact up to the relevant breakdown time scale $t \sim J^m/\lambda^{m+1}$ for each order $m$. We note that a small value of $\lambda$ is chosen to resolve the plateaus in $D$, which correspond to different orders in the DPT. However, the values of the 1st and 2nd order plateau are independent of $\lambda$ and depend only on the size and number of disconnected subspaces. Moreover, even at larger $\lambda$, when the plateaus are no longer separated, we find excellent agreement between DPT and exact evolution.

\subsection{Comparison of 3-body with XXZ model}\label{sec:3bxxz}

\begin{figure}[htb]
	\includegraphics[width=\linewidth]{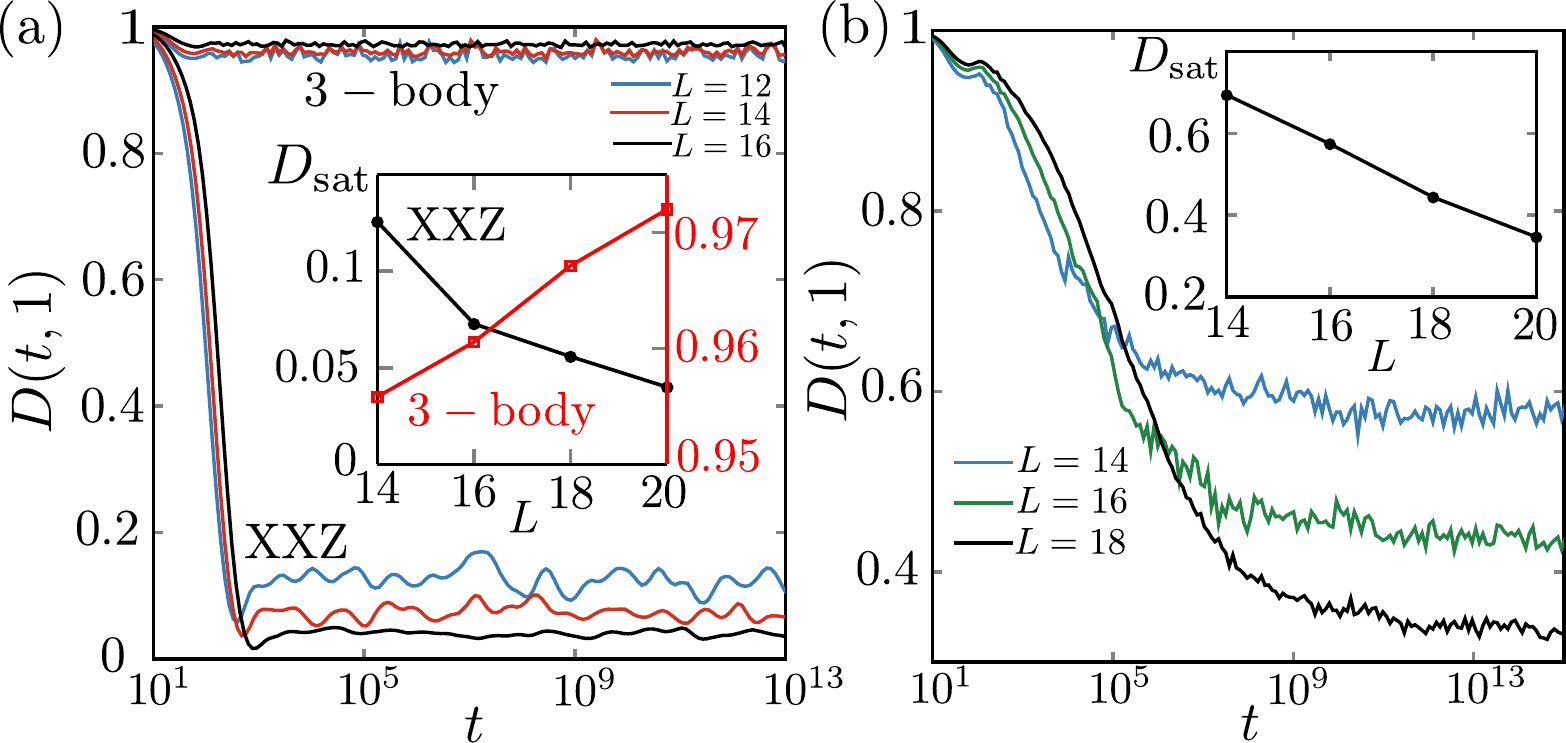}    	
  \vspace*{-7mm}
 \caption{ \label{fig:Sat} (Color online.) Evolution of the magnetization with momentum $k=1$ for various system sizes and $\lambda = 0.1$, in 1st order (a) and 2nd order (b) DPT. Plot (b) is for the 3-body model only. Insets show the system-size scaling of the saturation plateaus in DPT.
}
\end{figure}

After successfully benchmarking DPT against exact time evolution, we now use DPT to compare the relaxation in the 3-body model against the well-studied example of the XXZ chain which shows fast relaxation.  Fig.\ref{fig:Sat} shows $D(t,1)$ plotted for different sizes $L$ and orders of DPT. Notice that we do not terminate the evolution at the time scale  that would be relevant to each order, but we allow the system to evolve until it reaches a saturation plateau. This method allows us to measure how much the system relaxes in each order of DPT. It is obvious that including additional orders only lowers the values of saturation plateaus as each order contains the previous order plus some extra terms whose value is independent of the strength of the perturbation. 

Interestingly, the 1st order already reveals a clear difference between the 3-body and XXZ model, Fig.\ref{fig:Sat}(a). In the latter case, $D$ quickly decays to a small value ($\lesssim 0.1$), which further decreases with $L$ (inset). In the 3-body case, the plateau is close to 1 and grows with the system size $L$. This means the system does not relax on the time scales where only the 1st order is relevant, and is a direct consequence of the PCDSs in Fig.\ref{fig:TRGOP1}. As the system becomes larger, the more extended modes do not scatter, indicating that the fraction of extended non-local 1st order eigenstates vanishes. 

In the second order, Fig.\ref{fig:Sat}(b), we observe a completely different behaviour of the 3-body plateau which now  \emph{decreases} with $L$. Finite size scaling (inset) suggests that as one increases $L$, the system becomes progressively more delocalized in the 2nd order of DPT. We note, however, that delocalization in our DPT does not automatically imply ergodicity, since 
any order in DPT will have disconnected subspaces whose support is a vanishing fraction of the total Hilbert space. Once the model is in a delocalized regime, the entire Hilbert space can become connected by the action of unitary rotations, which generate subleading corrections to $D$.

\subsection{Relaxation time}

Using DPT we can furthermore scrutinize the possible similarity of our model in Eq.~(\ref{Ham}) with models showing ``quasi-MBL" behaviour~\cite{Yao14}. This can be done by investigating the finite-size scaling of the saturation time in DPT.



If the system delocalizes at a finite order in DPT, it is natural to expect that $D(t,k)$ obeys the following asymptotic behavior in time:
\begin{eqnarray}\label{eq:dpower}
D(t,k) \sim \exp(-A t^b),
\end{eqnarray}
where $A$ depends on momentum (and, therefore, system size $L$). On the other hand, if the system is ``quasi-MBL"~\cite{Yao14}, $D(t,k)$ should have the asymptotic form
\begin{eqnarray}\label{eq:dexp}
D(t,k) \sim \exp(-A \ln(t)),
\end{eqnarray}
which would yield a time scale for the relaxation of the smallest Fourier mode ($k=1$) that diverges exponentially with the size of the system.

\begin{figure}[htb]
\centering
\hspace*{-0.75cm}
\subfigure[]{%
\label{fig:first}%
\includegraphics[width=0.49\linewidth]{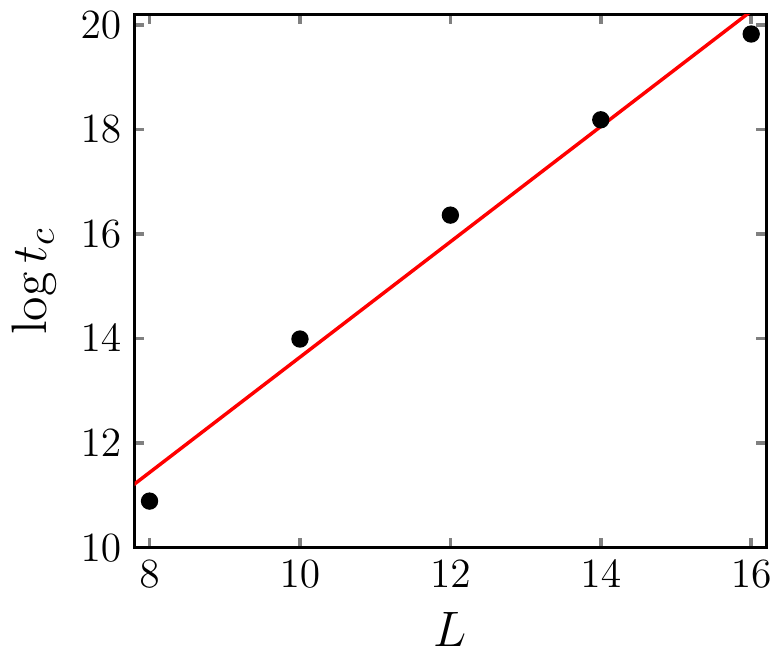}}%
\qquad
\hspace*{-0.75cm}
\subfigure[]{%
\label{fig:second}%
\includegraphics[width=0.49\linewidth]{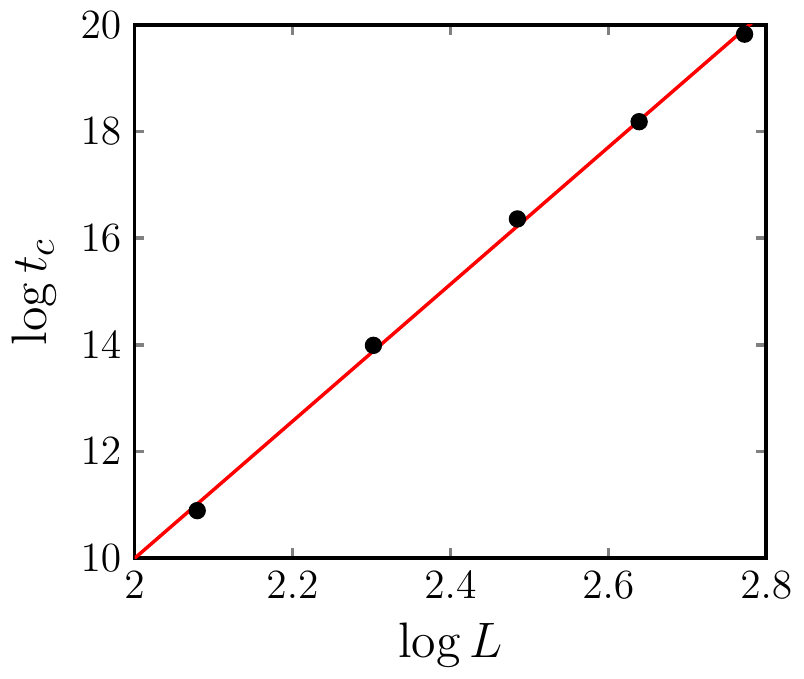}}%
\caption{\label{Fig:log_loglog} (Color online.) Scaling of delocalization time $t_c$ with system size $L$ for $\lambda=0.2$. Delocalization time is defined as the time it takes for magnetization to decay to 5\% of its initial value. Plotting the data on a single log-scale (a) and log-log scale (b) suggests that the magnetization of the 3-body model behaves according to Eq.~(\ref{eq:dpower}).}
\end{figure}

 In Fig.~\ref{Fig:log_loglog} we show the delocalization time (defined as the time it takes for magnetization to drop to 5\% of its initial value) as a function of system size $L$. This plot was obtained by exact diagonalization of the 3-body model for the fixed hopping $\lambda=0.2$.  Despite small system sizes, Fig.~\ref{Fig:log_loglog} suggests that Eq.~(\ref{eq:dpower}) is a much better fit to the data, suggesting that our 3-body model is in a different class from ``quasi-MBL" models.

 As an alternative way to probe the difference between our model in Eq.~(\ref{Ham}) and quasi-MBL models, we have also considered the standard quantifiers of integrability breaking used in random matrix theory, such as the statistics of energy level spacing. We found that the 3-body model is described by the Wigner-Dyson statistics already for $\lambda$ as small as 0.1. For this value of $\lambda$, we find the average value of the level statistics ``$r$" parameter~\cite{PalHuse} to be $\langle r \rangle \approx 0.53$ (this value is obtained for $L=20$ spins, after resolving translation and discrete symmetries of the model, and it was averaged over all eigenstates, which corresponds to the infinite temperature). The obtained value is very close to the Wigner-Dyson value $\langle r \rangle \approx 0.53$, and clearly inconsistent with the value expected for Poisson statistics, $\langle r \rangle\approx 0.39$. Since the level statistics probes the smallest energy scale in the system or,
equivalently, the asymptotically long time scales, this confirms our claim in Sec.~\ref{sec:3bxxz} that the 3-body model delocalizes at asymptotically long times, even though it displays slow dynamics over surprisingly long intermediate timescales.

\section{Generalization to other models}\label{sec:other}

Finally, we discuss some generalizations of the model in Eq.~(\ref{Ham}) in order to establish a more general understanding of the possible types of slow dynamics due to interaction constraints. We compare the polarization decay between different models from the point of view of 1st and 2nd order of DPT. The perturbation $V$ is always assumed to be the nearest neighbor (NN) hopping. The unperturbed Hamiltonian is chosen to have a combination of different terms, denoted by the following abbreviations: NNN stands for $\sum_{i}\sigma^{(3)}_{i}\sigma^{(3)}_{i+2}$, 3-body for $\sum_{i}\sigma^{(3)}_{i}\sigma^{(3)}_{i+1}\sigma^{(3)}_{i+2}$, 4-body  for $\sum_{i}\sigma^{(3)}_{i}\ldots\sigma^{(3)}_{i+3}$. ``Range-4" is used to denote all possible range-4 interactions. 

By combining these  interaction terms (with irrational coefficients, as mentioned in Sec.~\ref{sec:model}), various models can be constructed, and their dynamical behaviour (according to the behaviour of the plateau in 1st and 2nd order DPT) is summarized in Table I. For example,  we can observe that relaxation of the system is suppressed  up to order $m$ if $K-R \geq m$, where $K$,$R$ are the ranges of $H_{0}$ and $V$, respectively. However, this is a necessary condition but it is not sufficient. For example, by just adding 4-body interactions to the 3-body Hamiltonian, the system still delocalizes at 2nd order DPT. We believe that this condition becomes sufficient only if $H_{0}$ contains \emph{all possible} interactions up to that range. 
\begin{table}[htb] \label{eq:tab1}
\centering
 \begin{tabular}{| l | l | l | l |}
    \hline
    Model & 1st order & 2nd order \\ \hline
    XXZ & $\searrow $  & $\searrow $   \\ \hline
    Hopping+NNN & $\searrow$  & $\searrow $ \\ \hline
    Hopping+3-body & $\searrow$  & $\searrow $  \\ \hline 
    Hopping+NNN+3-body & $\searrow$  & $\searrow $  \\ \hline 
    XXZ+NNN & $\nearrow$ & $\searrow$   \\ \hline
    XXZ+3-body & $\nearrow$ & $\searrow$   \\ \hline
    XXZ+NNN+3-body & $\nearrow$ & $\searrow$  \\ \hline
    XXZ+NNN+3-body+4-body & $\nearrow$ & $\searrow$  \\ \hline
    XXZ+ all up to range-4 & $\nearrow$ & $\nearrow$ \\\hline
    \hline
    \end{tabular}
\caption{Behaviour of the saturation plateau of 1st and 2nd order of DPT as $L\to \infty$ for various models.   $\nearrow$ and $\searrow$ illustrate the fact that $D(t\to\infty)$ increases or decreases as a function of the system size.
}
\label{table:1}
\end{table}

\begin{figure}[h!]
\includegraphics[width=0.7\columnwidth]{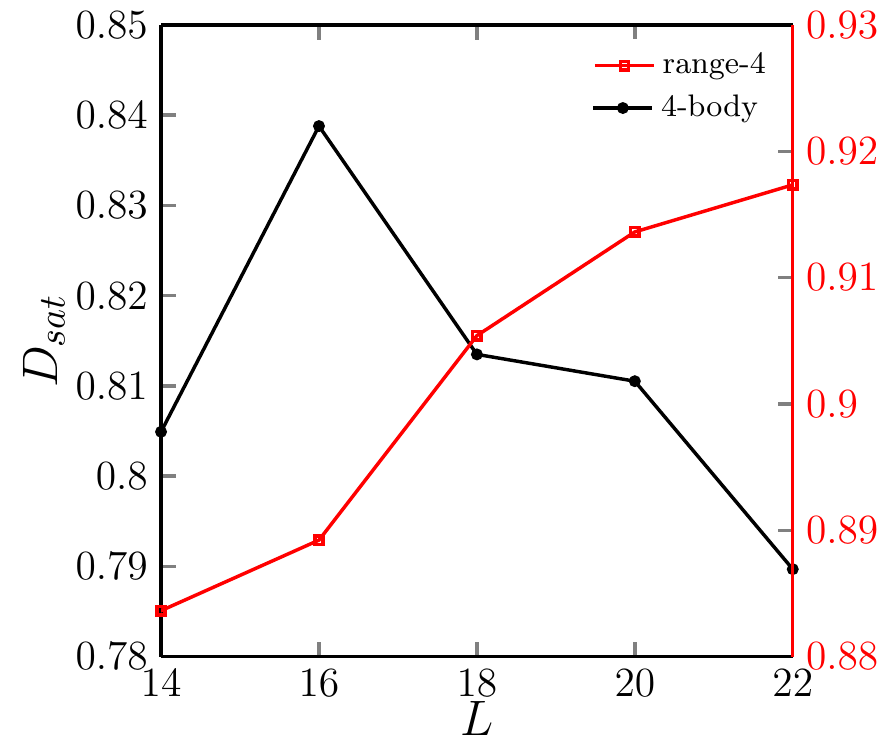}
\caption{ \label{Fig:4-body)} (Color online.) Scaling of the saturation plateaus of 3-body + 4-body model as well as the full range-4 model in 2nd order DPT. Adding 4-body interactions to the 3-body model is not enough to prevent relaxation  in the 2nd order. However, including all range-4 terms appears to prevent relaxation at that order.}
\end{figure} 

In order to corroborate the previous statement, we show that it is possible to prevent relaxation  in the second order of DPT, corresponding to the 2nd order plateau increasing with system size. For this, we require a Hamiltonian with 4-body interaction terms. Fig. \ref{Fig:4-body)} illustrates the saturation plateau of the 2nd order Hamiltonian for two different models where $K=4$, $R=2$. In the first case, a 4-body term ($\sum_{i}\sigma^{(3)}_{i}\ldots \sigma^{(3)}_{i+3}$)  is added to the 3-body Hamiltonian. In the second case, the most generic range-4 unperturbed Hamiltonian is chosen by adding all possible range-4 combinations of 2,3,4-body interactions. We observe that 4-body interactions in themselves are not enough to prevent relaxation of the system in the 2nd order as the saturation plateau decreases. On the other hand, the most generic range-4 interaction, obtained by taking the 3-body model and adding to it terms such as 
$ \sum_i \sigma^{(3)}_i \ldots \sigma^{(3)}_{i+3}$, 
$ \sum_i \sigma^{(3)}_i \sigma^{(3)}_{i+1}\sigma^{(3)}_{i+3}$, $ \sum_i \sigma^{(3)}_i \sigma^{(3)}_{i+2}\sigma^{(3)}_{i+3}$, 
$\sum_i \sigma^{(3)}_i \sigma^{(3)}_{i+3}$,
does indeed prevent relaxation in the 2nd order.

A general picture which emerges is that higher range terms in the diagonal part $H_0$ inhibit transport. More precisely, the previous results support our conjecture that a generic, range-$K$ translation-invariant interaction  leads to the absence of relaxation up to order $m=K-R$, where $R$ is the range of the hopping term $V$.  This is in line with the situation in MBL systems, where $H_0$ is expressed in terms of Local Integrals of Motion (LIOMs)~\cite{Serbyn13-1, Huse13} (LIOMs) and contains terms of arbitrary range (with decaying strengths). The LIOMs are expected to be robust to adding a small $V$, thus the system should stay localized in all orders of DPT. The DPT picture therefore presents a general framework which allows to understand truly localized systems, like disordered MBL models, as well as (local) translation-invariant systems that may display localization-like features only up to large but finite times.

\section{ Conclusion}\label{sec:conc}

We have introduced a general formalism to characterize slow dynamics in a broad class of systems with finite-range interactions and bounded local Hilbert space in any dimension. We illustrated the formalism and slow dynamics in a particular 1D model by demonstrating the plateaus in the decay of spin polarization and the log-like spreading of entanglement entropy. These results are insensitive to the choice of the parameters of the Hamiltonian as long as $\lambda \ll J$, i.e., they depend solely on the structure of the DPT expansion.

We showed that the dynamics can be significantly inhibited by changing the range of the diagonal term $H_0$. More precisely, the order $m$ plateau in the DPT approaches $1$ as $L\to\infty$ if all interaction terms with range $\leq m+2$ and with incommensurate amplitudes are included in $H_0$. Nevertheless, the system delocalizes by the order $m+1$ of DPT. Our 3-body model is an explicit example of this: it has a robust $m=1$ plateau, but delocalizes in $m=2$ order of DPT. Higher order plateaus, e.g., $m=2$, can be stabilized at the expense of including all interaction terms of range $\leq 4$. 

The general scenario of the absence of relaxation up to a finite time in translation-invariant systems should be contrasted with disordered MBL systems. 
In the latter case, $H_0$ is given in terms of LIOMs and contains interactions of arbitrary range with a decaying strength~\cite{Serbyn13-1, Huse13}. For small nonzero $\lambda$, the LIOMs are redefined and relaxation would be absent in all orders in DPT.
We also note that local models without a degenerate subspace exist \citep{Garrahan15}, where odd orders in DPT do not contribute and thus delay the onset of delocalization. Finally, it would be of interest to extend the DPT method to two component models~\cite{Muller, Yao14, QDL, QDL2, QDLEssler, Smith2017}, which generally become non-local when one particle species is integrated out. 

\section{Acknowledgements}

We thank Fran\c{c}ois Huveneers for useful discussions.
Z.P. and A.M. acknowledge support by EPSRC grant EP/P009409/1 and and the Royal Society
Research Grant RG160635. Statement of compliance with EPSRC policy framework on research data: This publication is theoretical work that does not require supporting research data.
D.A. acknowledges support by the Swiss National Science Foundation.
M.Z., M.M. and T.P. acknowledge Grants J1-7279 (M.Z.) and N1-0025 (M.M. and T.P.) of Slovenian Research Agency, and ERC grant No. 694544 - OMNES (T.P.).

\appendix

\section{Convergence with system size and the choice of the initial states}\label{sec:app1}

\begin{figure}[htb]
\includegraphics[width=0.8\linewidth]{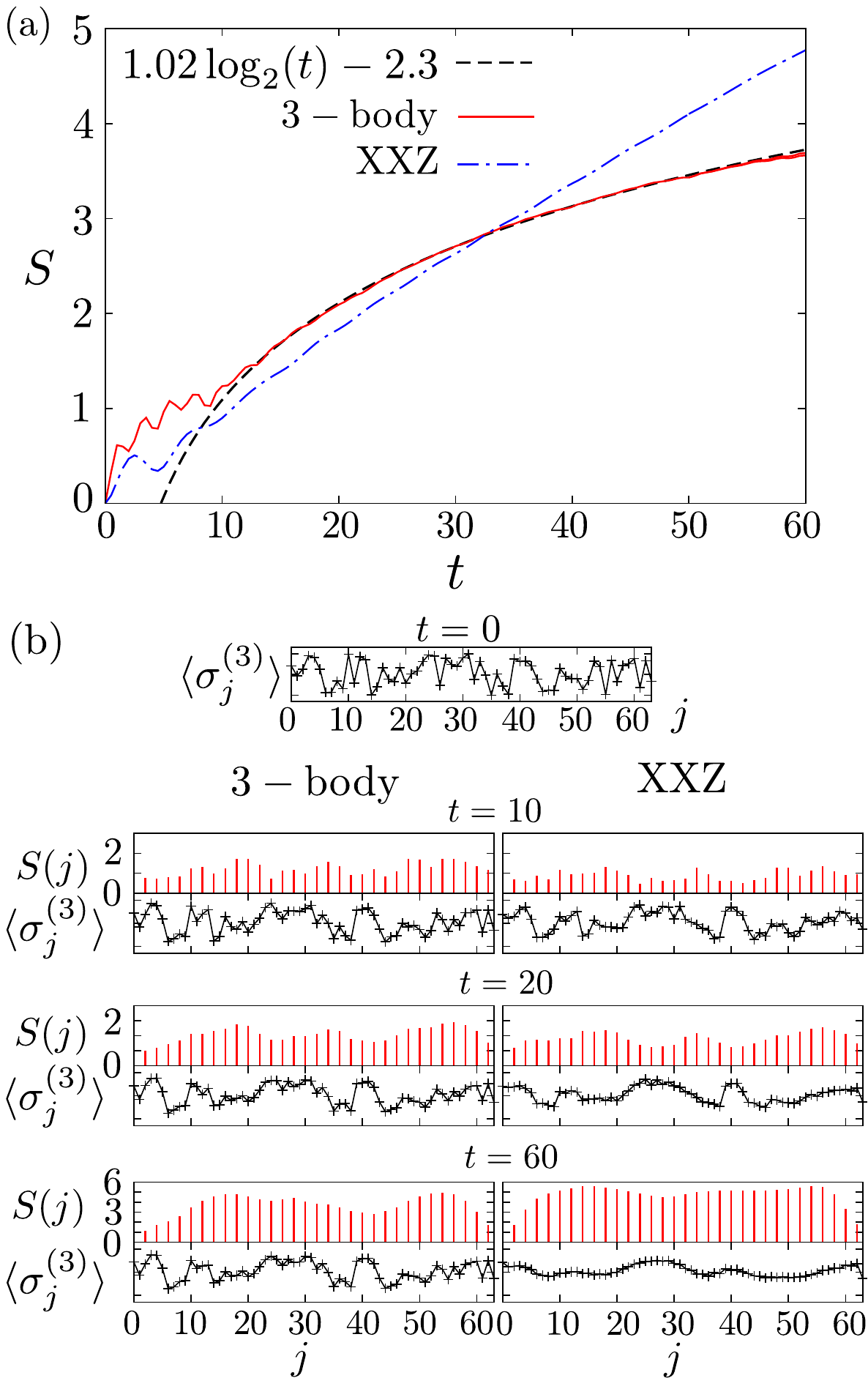}
\caption{Time evolution of the entanglement entropy for the initial product state where each spin is drawn uniformly on the Bloch sphere ($r=1$). (a) Slow growth of entropy in the 3-body model can be fitted by a logarithmic function of time, while the growth is linear in the XXZ model.  (b) Spatial profiles of magnetization and entanglement entropy at all cuts $j$ and at three different times. Note the larger entropy compared to Fig. 1 in the main text and therefore correspondingly shorter simulation times. All data is for $L=64$, $\lambda=0.2$.}
\label{fig:Bloch}
\end{figure}
Our results in the main text hold for generic initial product states. Two important special cases of such states are random computational states, i.e., states for which each spin is pointing either up or down (with equal probability), and random states in the sense of the Haar measure, i.e., states drawn uniformly on the Bloch sphere. In order to be able to smoothly interpolate between these two cases, we introduced a biased Bloch ensemble in Eq.~(\ref{eq:psi0}) parametrized by $r$ in Eq.~(\ref{eq:cos}).

State-to-state fluctuations in the entanglement entropy $S(t)$ are largest for $r=\infty$ and smallest for $r=1$ (uniform Bloch), while the average value of $S(t)$ is the largest for $r=1$ and smallest for $r=\infty$. The maximal time that can be simulated by the TEBD algorithm depends on  $S(t)$, and therefore large $r$ would be preferred. However, large $r$ would also necessitate a large ensemble size, in order to suppress state-to-state fluctuations. Therefore, some intermediate choice of $r$ would be optimal in practice. In the main text we have used $r=11$. We emphasize, however, that this choice is just for numerical convenience, and qualitatively similar results are obtained for other choices which we now demonstrate.

In Fig.~\ref{fig:Bloch} we show data similar to Fig.~\ref{Fig:big} in the main text, but here the initial state is unform on the Bloch sphere ($r=1$). One can see that even though the initial spins are not fully polarized, similar slow dynamics emerges as for the states with $r=11$. The difference is only in the prefactor of the log-like entropy growth, which is larger here, and thus only shorter times can be reached in the simulation. Spatial profile of the entanglement entropy  $S(j)$ for all cuts as well as magnetization profiles $\langle \psi(t)| \sigma_j^{(3)} |\psi(t)\rangle$ are again qualitatively different for the 3-body model compared to the XXZ chain.

\begin{figure}[htb]
\centerline{\includegraphics[width=0.8\linewidth]{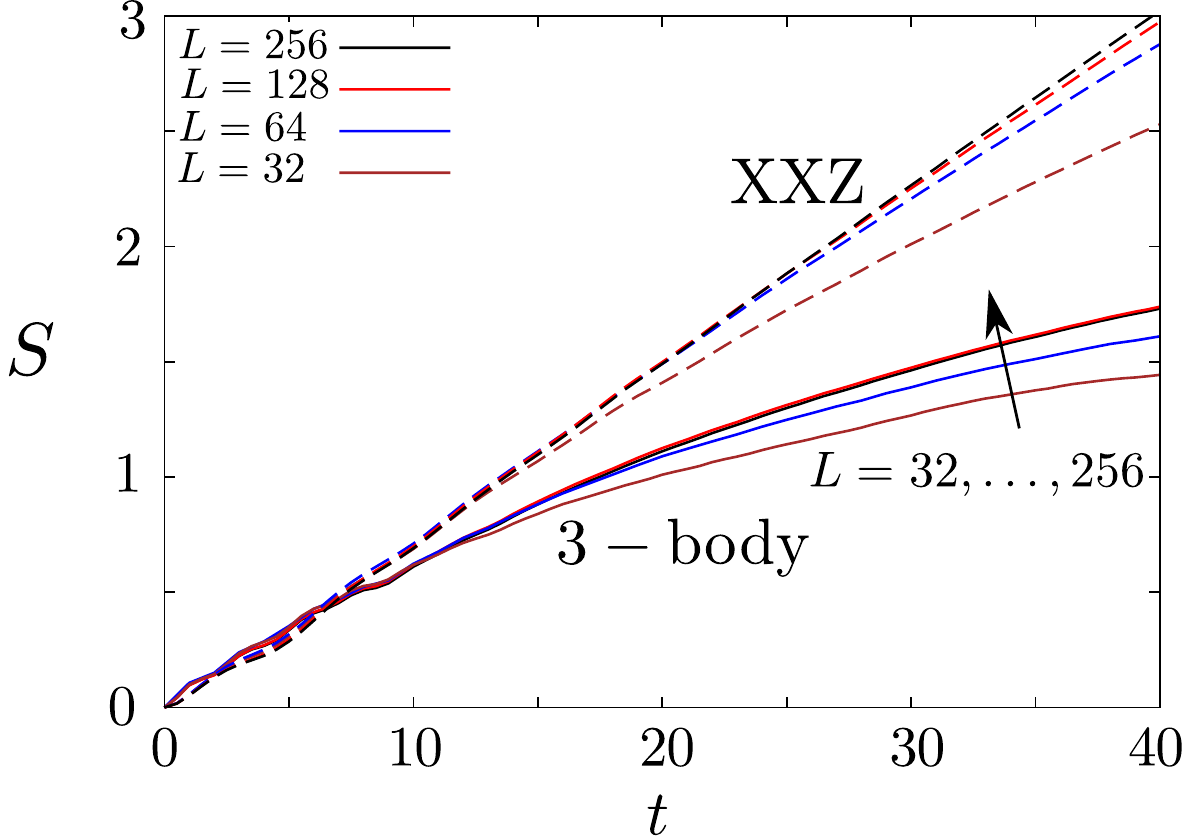}}
\caption{Entanglement entropy $S(t)$ (averaged over all bipartite cuts) for an ensemble of random product initial states with $r=11$ in Eq.~(\ref{eq:psi0}). Convergence with system size is achieved for $L\approx 128$ (for the 3-body model, $L=128$ and $L=256$ essentially overlap). Averaging is performed over $10-20$ initial states. All data is for $\lambda=0.2$.}
\label{fig:L}
\end{figure}
The initial state used in the main text is an instance of a state with $r=11$. This is the reason why initially spins are almost fully polarized in the $\pm z$-direction. In Fig.~\ref{fig:L} we show the \emph{average} entanglement entropy, where the averaging is done over the ensemble with $r=11$. The results shown in Fig.~\ref{fig:L} demonstrate that for the sizes used, the behavior becomes system-size independent.

\begin{figure}[htb]
\centerline{\includegraphics[width=0.8\linewidth]{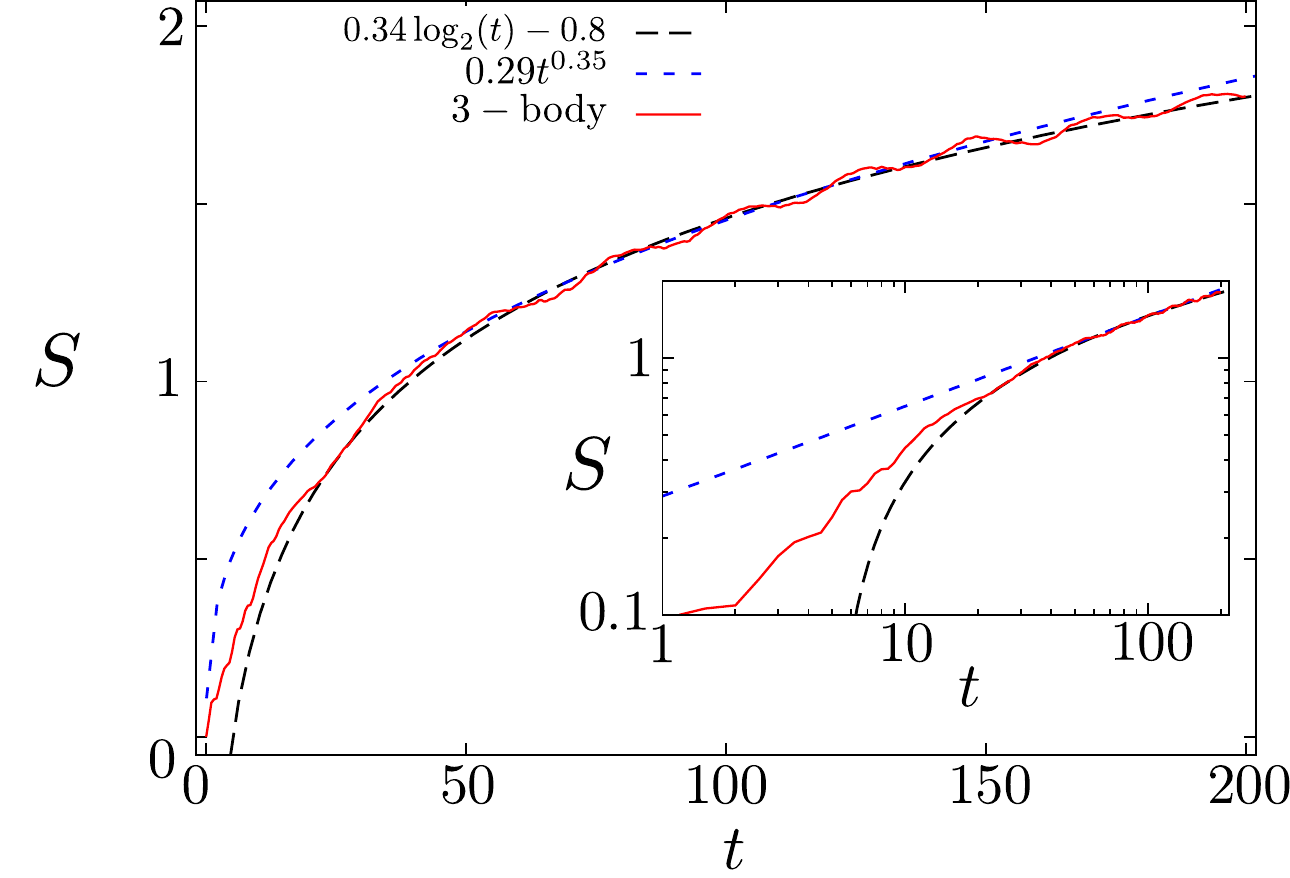}}
\caption{Entanglement entropy $S(t)$ (averaged over all bipartite cuts) for a single product initial state with $r=11$ used in Fig.~\ref{Fig:big} in the main text and the 3-body model. Logarithmic dependence (black) fits slightly better. Inset: log-log plot of the same data.}
\label{fig:fitS}
\end{figure}
Finally, we also illustrate the difficulty in numerically distinguishing the logarithmic dependence from a power law with a small power. In Fig.~\ref{fig:fitS}  we demonstrate that, while we can not exclude the possibility of a power-law growth of entanglement entropy in the 3-body model, logarithmic dependence appears to give better fit to the data.

\section{Application of degenerate perturbation theory to the 3-body model}\label{sec:dpt3b}

In this section we apply the general formalism of DPT in Sec.~\ref{sec:dpt} to the 3-body model in Eq.~(\ref{Ham}). This Hamiltonian is 3-local, leading to 6-local $F^{i}_{n_{1}n_{2}n_{3}}$'s:
\begin{eqnarray}\label{ap:generators}
\nonumber F^{i}_{-400} &=&(\mathds{1} + \Pi_{i})[ p_{i-2}p_{i-1}\sigma^{-}_{i}\sigma^{+}_{i+1}q_{i+2}q_{i+3}],\\
\nonumber F^{i}_{-44-4} &=& (\mathds{1} + \Pi_{i})[p_{i-2}p_{i-1}\sigma^{-}_{i}\sigma^{+}_{i+1}q_{i+2}p_{i+3}],\\
\nonumber F^{i}_{-444} &=&  (\mathds{1} + \Pi_{i})[q_{i-2}p_{i-1}\sigma^{-}_{i}\sigma^{+}_{i+1}q_{i+2}q_{i+3}],\\
\nonumber F^{i}_{-480} &=&  (\mathds{1} + \Pi_{i})[q_{i-2}p_{i-1}\sigma^{-}_{i}\sigma^{+}_{i+1}q_{i+2}p_{i+3}],\\
\nonumber F^{i}_{0-4-4} &=&  (\mathds{1} + \Pi_{i})[q_{i-2}p_{i-1}\sigma^{+}_{i}\sigma^{-}_{i+1}p_{i+2}p_{i+2}],\\
\nonumber F^{i}_{0-44} &=&  (\mathds{1} + \Pi_{i})[q_{i-2}q_{i-1}\sigma^{+}_{i}\sigma^{-}_{i+1}q_{i+2}p_{i+3}],\\
\nonumber F^{i}_{000} &=&  (\mathds{1} + \Pi_{i})[p_{i-2}q_{i-1}\sigma^{-}_{i}\sigma^{+}_{i+1}q_{i+2}p_{i+3}+\\
\nonumber &+& q_{i-2}q_{i-1}\sigma^{-}_{i}\sigma^{+}_{i+1}q_{i+2}q_{i+3}+\\
 &+& p_{i-2}p_{i-1}\sigma^{-}_{i}\sigma^{+}_{i+1}p_{i+2}p_{i+3}+\\
\nonumber &+& q_{i-2}p_{i-1}\sigma^{-}_{i}\sigma^{+}_{i+1}p_{i+2}q_{i+3}],\\
\nonumber F^{i}_{04-4} &=& (\mathds{1} + \Pi_{i})[q_{i-2}q_{i-1}\sigma^{-}_{i}\sigma^{+}_{i+1}q_{i+2}p_{i+3}],\\
\nonumber F^{i}_{044} &=& (\mathds{1} + \Pi_{i})[q_{i-2}p_{i-1}\sigma^{-}_{i}\sigma^{+}_{i+1}p_{i+2}p_{i+3}],\\
\nonumber F^{i}_{4-80} &=& (\mathds{1} + \Pi_{i})[q_{i-2}p_{i-1}\sigma^{+}_{i}\sigma^{-}_{i+1}q_{i+2}p_{i+3}],\\
\nonumber F^{i}_{4-4-4} &=& (\mathds{1} + \Pi_{i})[q_{i-2}p_{i-1}\sigma^{+}_{i}\sigma^{-}_{i+1}q_{i+2}q_{i+3}],\\
\nonumber F^{i}_{4-44} &=& (\mathds{1} + \Pi_{i})[p_{i-2}p_{i-1}\sigma^{+}_{i}\sigma^{-}_{i+1}q_{i+2}p_{i+3}],\\
\nonumber F^{i}_{400} &=& (\mathds{1} + \Pi_{i})[p_{i-2}p_{i-1}\sigma^{+}_{i}\sigma^{-}_{i+1}q_{i+2}q_{i+3}],
\end{eqnarray}
where $n_{1}$, $n_{2}$, $n_{3}$ are associated with the nearest-neighbour (NN), next-nearest-neighbour (NNN), and 3-body interactions, respectively, while $p_{i}$, $q_{i}$ are the projectors to the  $\uparrow$, $\downarrow$ spin on the $i$th site, i.e., $p_{i} = \text{diag}(1,0)$, $q_{i} = \text{diag}(0,1)$. The operator $\Pi_{i}$ performs a reflection of an operator around the bond $i$, i.e., $\Pi_{i}(O_{i}O_{i+2}) = O_{i+1}O_{i-1}$. The reflection symmetry of $F$'s is a consequence of the reflection symmetry of the full Hamiltonian. 

\bibliography{3bodyNotes}
\bibliographystyle{apsrev4-1}

\end{document}